\documentclass[amsmath,amssymb,aps,graphicx,twocolumn,reprint,prl]{revtex4-2}
\usepackage{graphicx}
\usepackage{physics}
\usepackage{color}
\usepackage{mathrsfs}
\usepackage[caption=false]{subfig}
\begin{document}
\newcommand{\dif}{\mathrm{d}}

\title{Frequency-tunable microwave quantum light source based on superconducting quantum circuits} 

\author{Yan Li$^{1}$}
\altaffiliation{These three authors contributed equally to this work.}

\author{Zhiling Wang$^{1}$}
\altaffiliation{These three authors contributed equally to this work.}

\author{Zenghui Bao$^{1}$}
\altaffiliation{These three authors contributed equally to this work.}

\author{Yukai Wu$^{1,2}$}

\author{Jiahui Wang$^{1}$}

\author{Jize Yang$^{1}$}

\author{Haonan Xiong$^{1}$}

\author{Yipu Song$^{1,2}$}

\author{Hongyi Zhang$^{1,2}$}
\email{hyzhang2016@tsinghua.edu.cn}

\author{Luming Duan$^{1,2}$}
\email{lmduan@tsinghua.edu.cn}

\affiliation{$^{1}$Center for Quantum Information, Institute for Interdisciplinary Information Sciences, Tsinghua University, Beijing 100084, PR China}

\affiliation{$^{2}$Hefei National Laboratory, Hefei 230088, PR China}

\date{March 15, 2023}

\begin{abstract}

A nonclassical light source is essential for implementing a wide range of quantum information processing protocols, including quantum computing, networking, communication, and metrology. In the microwave regime, propagating photonic qubits that transfer quantum information between multiple superconducting quantum chips serve as building blocks of large-scale quantum computers. In this context, spectral control of propagating single photons is crucial for interfacing different quantum nodes with varied frequencies and bandwidth. Here we demonstrate a microwave quantum light source based on superconducting quantum circuits that can generate propagating single photons, time-bin encoded photonic qubits and qudits. In particular, the frequency of the emitted photons can be tuned in situ as large as 200 MHz. Even though the internal quantum efficiency of the light source is sensitive to the working frequency, we show that the fidelity of the propagating photonic qubit can be well preserved with the time-bin encoding scheme. Our work thus demonstrates a versatile approach to realizing a practical quantum light source for future distributed quantum computing. 

\end{abstract}

\maketitle 

\section*{Introduction}

Photons in non-classical states are of great importance for both fundamental research and quantum applications. They provide major testbeds for quantum optics experiments~\cite{fabre_modes_2020,gu_microwave_2017}. In the meantime, photons serve as the backbone of quantum networks~\cite{kimble_quantum_2008}, which enables various quantum technologies including quantum communication~\cite{BB84, RevModPhys.92.025002} and large-scale quantum computation~\cite{reiserer_cavity_based_2015,Monroe2010}.
Photons in squeezed states, which can be generated through the parametric conversion process, are important tools for quantum metrology~\cite{bienfait_magnetic_2017,braunstein_quantum_2005} and measurement-based quantum computing~\cite{asavanant_generation_2019,larsen_deterministic_2019}. Other than that, as an ideal carrier of quantum information, single photon sources are widely used for entanglement distribution and remote entanglement generation~\cite{Cirac_97_transfer,kimble_quantum_2008}. In the microwave regime, single photons and propagating photonic qubits can be used to coherently link remote quantum chips to realize modular quantum computing and scale up computing power for superconducting quantum devices~\cite{Devoret2016,Kurpiers_2018,Kurpiers_2019,zhong2021,zhong2023}. A quantum light source that generates those quantum states of photons is thus a fundamental device in the field of quantum information processing~\cite{braunstein_quantum_2005,gu_microwave_2017}.

Single photons in the optical frequency can be generated with atoms, solid-state quantum dots, or through parametric down-conversion in nonlinear crystals~\cite{shields_semiconductor_2007,senellart_high_performance_2017,Thomas_rempe_2022}. In the microwave regime, matter qubits can strongly couple to microwave cavities, which may facilitate the photon generation process~\cite{blais_circuit_2021,gu_microwave_2017,Bao_2022,Wang_2022}. Superconducting qubits can be used as ancilla to generate the specific quantum state of photons through either spontaneous emission~\cite{eichler_experimental_2011,lang_correlations_2013,besse_realizing_2020} or cavity-assisted Raman-type transition~\cite{pechal_microwave_controlled_2014,Kurpiers_2019,ilves_demand_2020}.
For the latter case, the pulse shape of the emitted photons can be conveniently controlled by shaping the driving pulse. Indeed, spectral control of the quantum light source is of great importance in many circumstances. For instance, when the propagating photons are used for quantum state transfer and entanglement generation, a frequency-tunable quantum light source is necessary to bridge the possible spectral mismatches of distant quantum nodes~\cite{zhu_spectral_2022}. Apart from that, for achieving high transfer fidelity and efficiency, a well-controlled bandwidth and pulse shape of the photon field is required to fulfill temporal mode matching among different nodes~\cite{Cirac_97_transfer,Klaus_2019}.

In this work, we demonstrate a frequency-tunable microwave single-photon source based on superconducting quantum circuits. The single photon source can also be used to generate various quantum states of propagating photons, including time-bin encoded photonic qubits and qudits. We further showcase that by encoding the photonic quantum state in the Fock-1 subspace, the fidelity of the prepared photon states is insensitive to the inevitable circuit loss and non-unity quantum efficiency of the light source, which is particularly important in the microwave regime. As such, our work provides an important building block for future modular quantum computing and microwave quantum optics.

\section*{the device}

\begin{figure}[!tbp]
\centering
\includegraphics[width=0.9\linewidth]{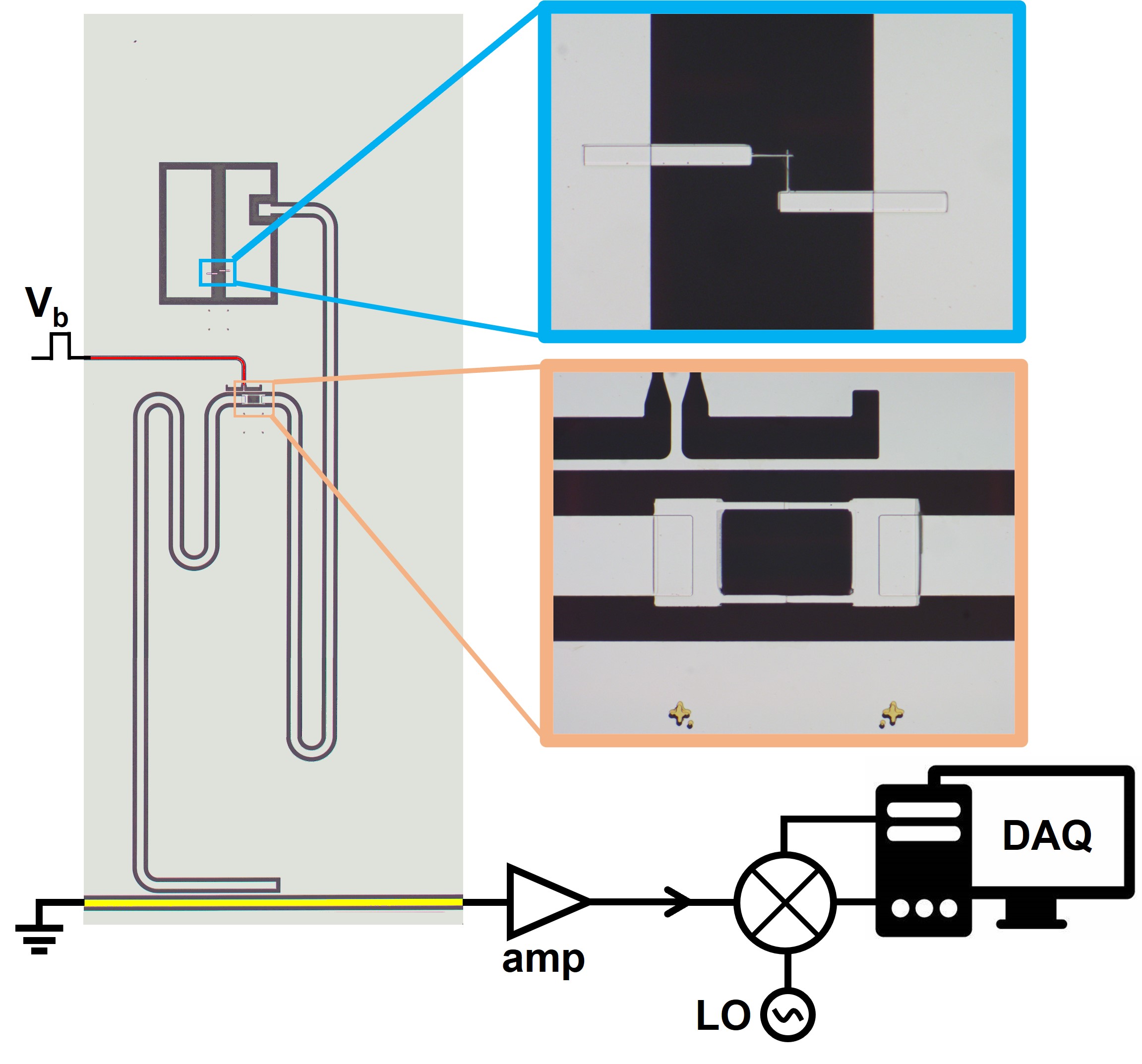}
\caption{\textbf{An illustration of the device and measurement setup.} An optical microscope image of the microwave quantum light source and a schematic of the homodyne setup for the characterization of the propagating photon field. The upper inset shows a zoom shot of the Josephson junction of the transmon qubit. The lower inset is a zoom shot of the superconducting quantum interference device used for frequency adjustment of the photon source. }
\label{device}
\end{figure}

As illustrated in Fig.~\ref{device}, the device composes a $\lambda$/2 coplanar waveguide resonator dispersively coupled to a transmon qubit. A superconducting quantum interference device (SQUID) is embedded in the middle of the resonator to tune the resonator frequency~\cite{Bao_memory_2021}. 
The transmon qubit is used to facilitate the generation of the quantum state of light. The transition frequency between the ground state $\ket{g}$ and the first exited state $\ket{e}$ of the transmon is 6.1362~GHz, with a charging energy $E_c$=217~MHz. The relaxation time for the first and second excited state is $T_1^{ge}=8.89~\mu s$ and $T_1^{ef}=6.99~\mu s$, respectively. The dephasing time is measured to be $T_2^{ge}=2.88~\mu s$ and $T_2^{ef}=3.16~\mu s$. In general, the excited-state population of the qubit can be transferred to photons in the resonator via either spontaneous decay~\cite{Wallraff2011,lang_correlations_2013} or the resonator-assisted Raman transition~\cite{pechal_microwave_controlled_2014}, after which the photons escape from the resonator to the transmission line as propagating photons.

The SQUID used for tuning the resonator frequency consists of two parallel Josephson junctions, which behave as a tunable inductor $L_j$ by varying the flux $\varphi$ through the loop of the Josephson junctions. 
The total inductance of the SQUID embedded resonator is determined by a sum of the resonator inductance $L_0$ and the flux-dependent SQUID inductance $L_j(\varphi)$. Correspondingly, the resonance frequency can be written as $\omega=\sqrt{1/(L_0+L_j(\varphi))C_0}$, where $C_0$ is the resonator capacitance and the capacitance of SQUID is small enough to be neglected. Therefore, it is possible to tune the resonator frequency via an external magnetic field~\cite{Sandberg_Tuning_2008,Bao_memory_2021}. In our device, the magnetic field is introduced by applying an analog voltage on the flux line, as shown in Fig.~\ref{device}, which is linearly related with the magnetic flux seen by the SQUID loop.

The sample is cooled down to about 20~mK in a dilution fridge. As illustrated in Fig.~\ref{device}, the microwave photons emitted from the quantum light source are first amplified with a Josephson parametric amplifier, and then analyzed with a homodyne setup, where the quadratures of the output photon field are recorded. From the measured quadratures, we can calculate moments of the propagating photon field, and the Winger function or density matrix can be correspondingly reconstructed. Details about this method can be found in Ref.~\cite{Eichler_Characterizing_2012}.

\section*{frequency-tunable single photon source}

\begin{figure}[!tbp]
\centering
\includegraphics[width=1\linewidth]{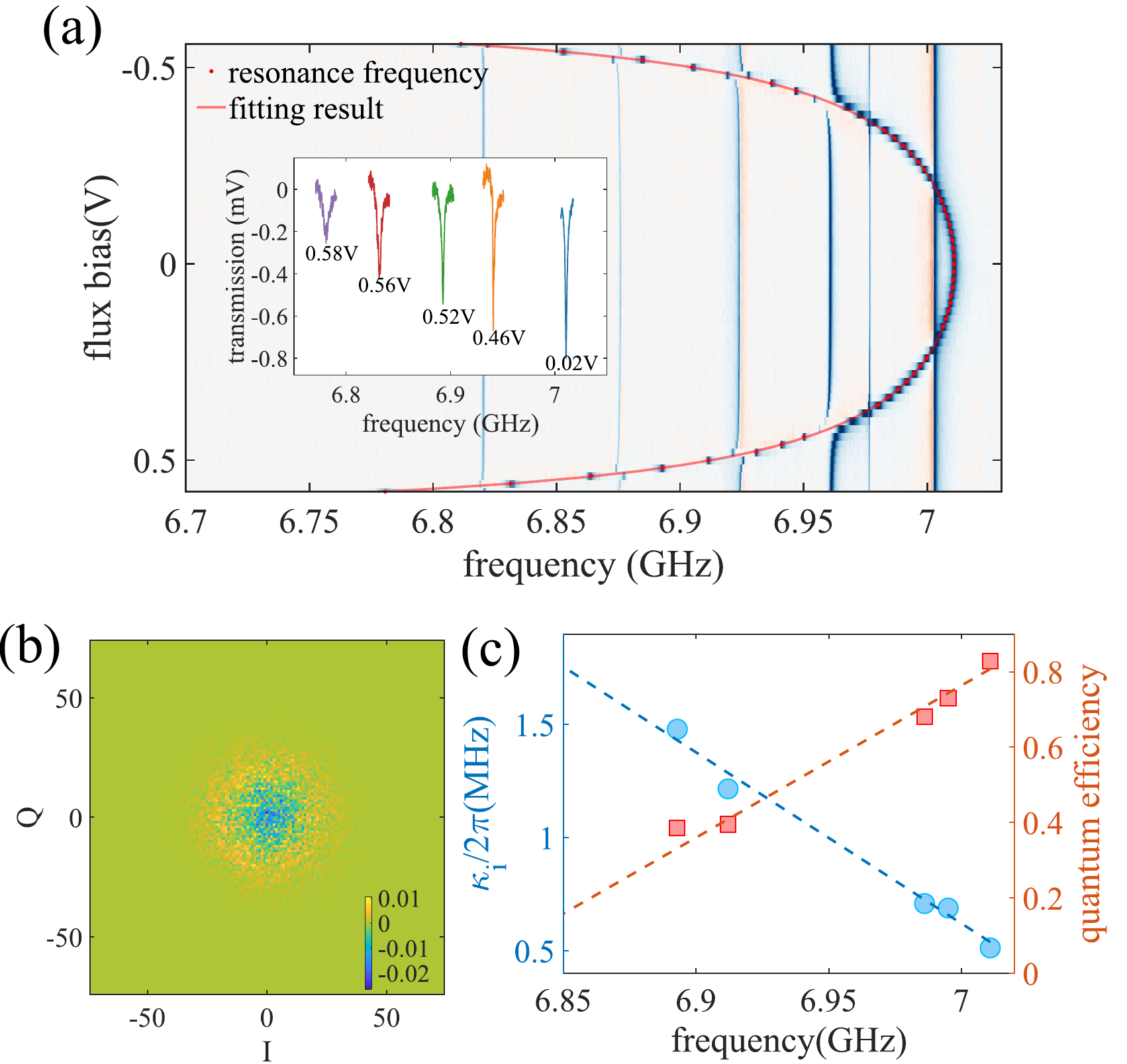}
\caption{\textbf{Frequency-tunable single photon source.} (a) A color plot of the transmission spectra of the frequency-tunable resonator with varied flux bias. The transmission dips that show minor flux dependence originate from other resonators on the same chip. The inset of (a) shows some typical transmission spectra of the resonator under varied flux bias, from where one could see that the resonance frequency can be tuned by over 200~MHz. (b) Histogram of the in-phase and out-of-phase quadratures of the single photon field, which is obtained by subtracting the quadrature distribution of a vacuum field from that when the single photons are prepared. (c) The internal loss rate $\kappa_i$ (blue circles) and the measured quantum efficiency (red squares) of the photon source show a linear dependence on the frequency of the photon source. The dashed lines are linear fits of the corresponding data.}
\label{SP}
\end{figure}

We use the resonator-assisted Raman process to generate propagating single photons, wherein the qubit population in the second excited state $\ket{f}$ is transferred as a single photon in the resonator via the `f0g1' transition. It has been shown that by using `f0g1' driving pulse with different shapes, lengths and phases, the pulse shape, bandwidth and phase of the single photons can be well-controlled~\cite{pechal_microwave_controlled_2014}, which promises that such a single photon source can be used in scenarios with varied requirements on the properties of single-photon pulses~\cite{Lukin2022,zhu_spectral_2022}.

Fig.~\ref{SP}(b) shows the histogram of the in-phase and out-of-phase quadratures of the propagating single photon field, by subtracting the measured quadrature histogram of the vacuum field from that when the photon is released from the resonator. One could see a clear signature of the single photon state, with a circular distribution in the phase space~\cite{eichler_experimental_2011}. From the measured quadrature histograms, we determine the average photon number of the propagating field to be 0.414. Considering that a hanger resonator is used for the single photon generation (see Fig.~\ref{device}), only half of the photons emitted to the transmission line are sent to the homodyne detector. We estimate that the quantum efficiency of the single photon source is 82.8$\%$. The inefficiency can be mainly attributed to the internal loss of the resonator, which is estimated to be $\kappa_i$/($\kappa_c+\kappa_i$), where $\kappa_i$ and $\kappa_c$ represents the internal loss rate and the out-coupling rate of the resonator, respectively. By fitting the measured transmission spectra, we have $\kappa_i$/($\kappa_c+\kappa_i$) = 17$\%$, with $\kappa_i/2\pi$ = 0.51 MHz and $\kappa_c/2\pi$ = 2.49 MHz, which agrees well with the estimated quantum efficiency of the photon source.

As discussed above, the frequency of the emitted photon is determined by the resonator frequency, which can be tuned by varying bias voltage applied to the flux line.
In Fig.~\ref{SP}(a) we measure the transmission spectra of the resonator as a function of the applied voltage on the flux line with a vector network analyzer, where the resonance frequency can be continuously tuned by more than 200 MHz. It is worth noting that when the resonance frequency is tuned away from the sweep spot, the dip of the transmission spectra becomes smaller, as shown in the inset of Fig.~\ref{SP}(a), which indicates an increasing internal loss rate. In Fig.~\ref{SP}(c) we present the fitted $\kappa_i$ as a function of the resonator frequency, where $\kappa_i/2\pi$ increases from 0.51~MHz at the sweet spot to 1.48~MHz when the resonance frequency is shifted by about 120~MHz. Similar phenomena have been reported in previous works~\cite{Sandberg_Tuning_2008}, which is attributed to energy dissipation by the subgap resistance of the SQUID.

The increased internal loss rate would degrade the quantum efficiency of the single-photon source. In Fig.~\ref{SP}(c), we measure average photon numbers of the emitted photon field with varied photon frequencies. As expected, tuning the single-photon frequency away from the sweet spot by about 120~MHz results in a degradation of the quantum efficiency from 82.8$\%$ to 38.6$\%$, which is mainly caused by the resonator internal loss-induced inefficiency of 50.5$\%$ (with $\kappa_i/2\pi$ = 1.48 MHz and $\kappa_c/2\pi$ = 1.45 MHz). In addition, we observe decreased energy relaxation times (T1) for both $\ket{e}$ state and $\ket{f}$ state of the transmon qubit when the resonator frequency is detuned, which may also contribute to the decreased quantum efficiency. Such a limited quantum efficiency of the single-photon source would result in poor fidelity when directly using the single-photon state as propagating qubits.

\section*{Propagating photonic qubit}

\begin{figure}[!tbp]
\centering
\includegraphics[width=1\linewidth]{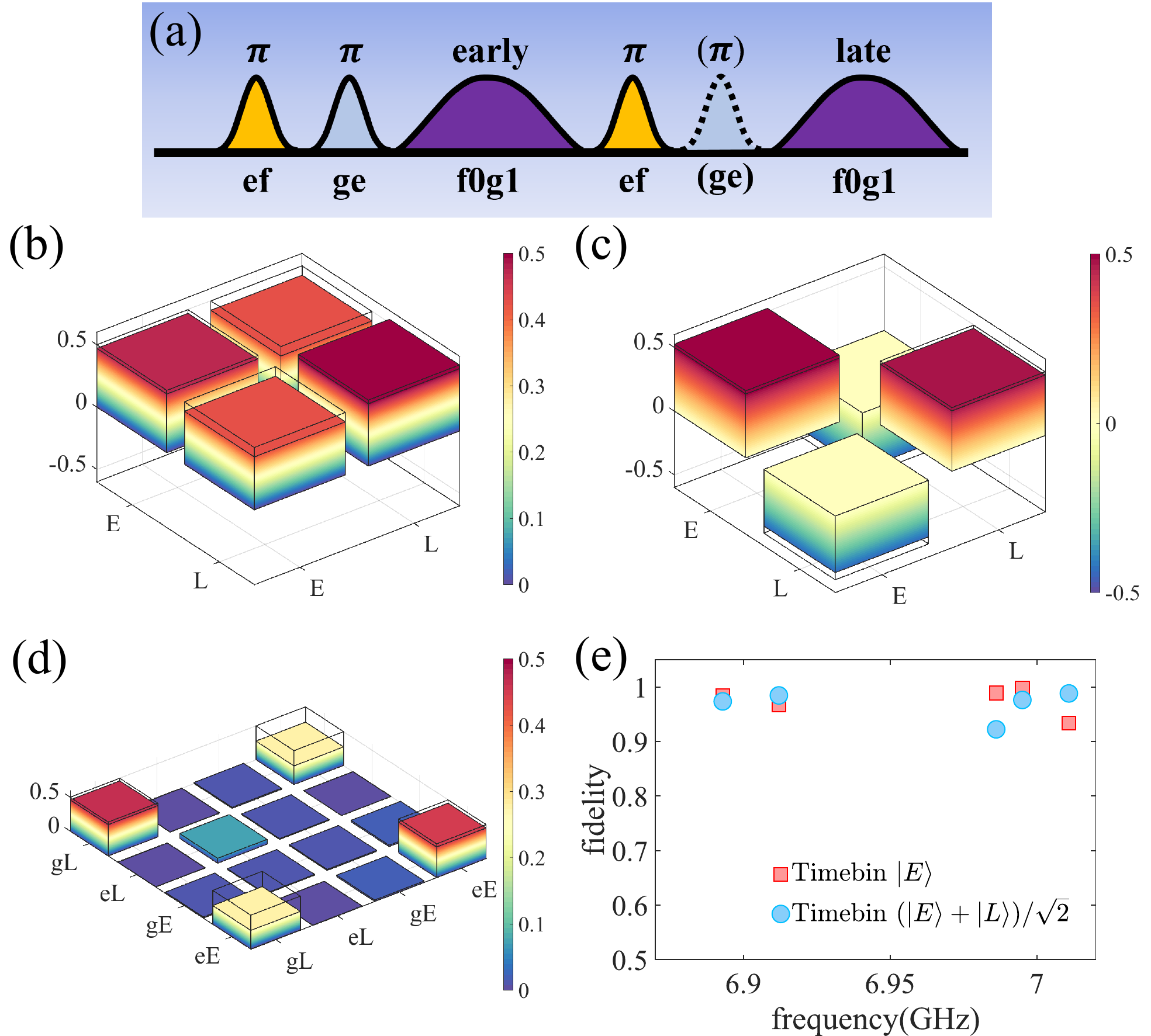}

\caption{\textbf{Frequency-tunable photonic qubit.} (a) Pulse sequence for the generation of time-bin encoded photonic qubit. If the qubit state is firstly prepared in a superposition state $\cos{\frac{\theta}{2}}\ket{g}+\sin{\frac{\theta}{2}}\ket{e}$, after the pulse sequence we would have $\cos{\frac{\theta}{2}}\ket{L}+\sin{\frac{\theta}{2}}\ket{E}$. The second $\pi_{ge}$ pulse (dashed envelope) is required to generate entanglement between the transmon qubit and the photonic qubit. The $\pi_{ge}$ and $\pi_{ef}$ drives are derivative removal by adiabatic gate pulses with a length of 80 ns. The ‘f0g1’ pulse has a peak-truncated Gaussian envelope and a duration of 400 ns. (b) The reconstructed density matrice for a time-bin qubit in $(\ket{E}+\ket{L})/\sqrt{2}$, and in (c) $(\ket{E}-\ket{L})/\sqrt{2}$. (d) The reconstructed density matrix for the qubit-photon entangled state $(\ket{g}\ket{L}+\ket{e}\ket{E}/\sqrt{2})$  with a fidelity of 73.84$\%\pm$0.42$\%$. The corresponding ideal matrix elements are indicated with the hollow caps. (e) The measured state fidelities for the time-bin qubits $\ket{E}$ and $(\ket{E}+\ket{L})/\sqrt{2}$ with varied frequencies, which shows minor dependence on the photon frequency.}
\label{time-bin}
\end{figure}

In the presence of inefficiency of the frequency-tunable single-photon source, time-bin encoding can be used to realize propagating photonic qubits with high fidelity. The qubit state is defined as a single photon that exists in a superposition of two temporal modes~\cite{ilves_demand_2020}. The photon that exists in the early or late temporal mode is written as $\ket{E} \equiv \ket{10}$ or $\ket{L} \equiv \ket{01}$, where $\ket{10}$ ($\ket{01}$) represents that the single photon is in the early (late) mode. Since the qubit state is defined in the single-photon subspace, the photon-loss-induced error would leave the photonic state out of the qubit space, and thus can be excluded by projecting the system state to the single-photon subspace~\cite{Kurpiers_2019,ilves_demand_2020}. This is very important for propagating microwave photonic qubits considering the strong propagation loss and insertion loss for photons in the microwave regime. Moreover, it has been demonstrated that the time-bin qubit is robust to phase noise, which makes them an excellent candidate as carriers of quantum information for quantum state transfer and entanglement distribution in microwave quantum networks~\cite{ilves_demand_2020}.

The pulse sequence for the generation of the time-bin qubit can be found in Fig.~\ref{time-bin}(a). If the transmon qubit of the photon source is prepared in a superposition state $\alpha\ket{g}+\beta\ket{e}$, after the pulse sequence one would have $\alpha\ket{L}+\beta\ket{E}$, wherein the quantum state defined with the transmon is transferred to the photonic qubit. In this way, we prepare photonic qubits in different quantum states when the resonator frequency is placed at the sweet spot, and some typical density matrices can be found in Fig.~\ref{time-bin}(b) and (c). It is worth noting that the density matrices are obtained by projecting the reconstructed two-mode photonic state to the single-photon subspace~\cite{Eichler_Characterizing_2012}. Consequently, even though the internal quantum efficiency of the photon source is not unity, the fidelity of the photonic qubit can still be above 92.2$\%$ thanks to the time-bin encoding scheme. In such cases, the photonic qubit fidelity is mainly bottlenecked by the limited energy relaxation rate and dephasing rate of the transmon qubit.

We note that quantum entanglement between the transmon qubit and the propagating photonic qubit can be generated. With the pulse sequence shown in Fig.~\ref{time-bin}(a), we prepare the system into a Bell state $(\ket{g}\ket{L}+\ket{e}\ket{E}/\sqrt{2})$ by initially preparing the transmon in a superposition of $(\ket{g}+\ket{e})/\sqrt{2}$. The reconstructed density matrix is shown in Fig.~\ref{time-bin}(d), with a fidelity of 73.84$\% \pm$0.42$\%$. Such a qubit-photon entanglement is a crucial resource for the entanglement generation and distribution among remote superconducting qubits.

The frequency of the propagating time-bin qubit can be adjusted by tuning the resonance frequency of the resonator. In the experiment, we measure the fidelity of the time-bin qubit at different frequencies, as presented in Fig.~\ref{time-bin}(e). One could see that the fidelity of both $\ket{E}$ state and $(\ket{E}+\ket{L})/\sqrt(2)$ state show minor dependence on the photon frequency. As discussed above, the time-bin encoding scheme ensures that the fidelity of the propagating qubit is insensitive to the quantum efficiency of the light source. Even though the quantum efficiency of the photon source drops from 82.8$\%$ to 38.6$\%$ with a detuned photon frequency, the fidelity of the time-bin can still keep well above 90$\%$. This result promises that the single-photon source can be used to align to spectrally mismatched quantum nodes with high fidelity.

\section*{Propagating photonic qutrit}

\begin{figure}[!tbp]
\centering
\includegraphics[width=1\linewidth]{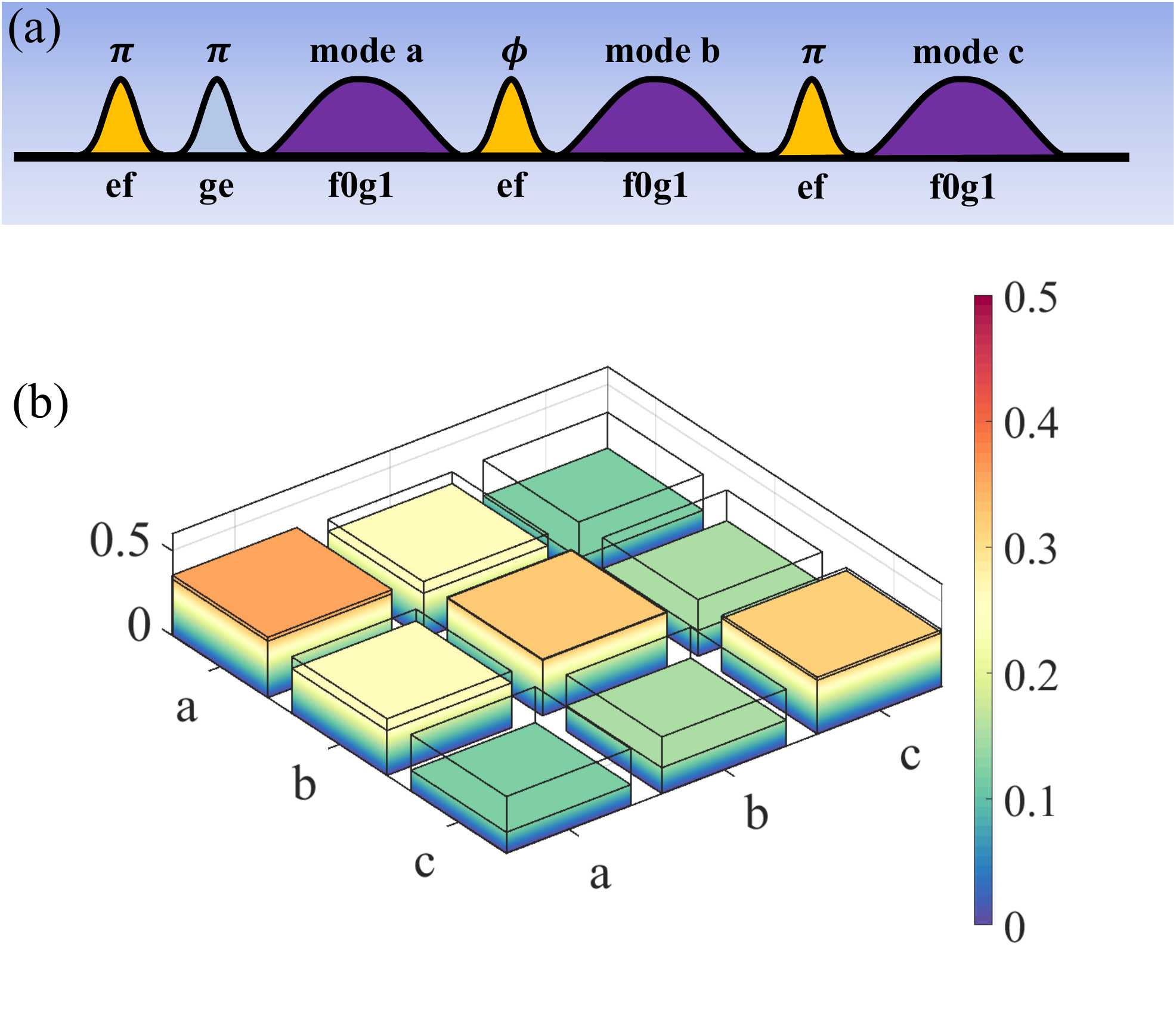}
\caption{\textbf{Photonic qutrit states.} (a) The pulse sequence to generate propagating photonic qutrit with the quantum light source. The qubit state is firstly prepared in a superposition state $\cos{\frac{\theta}{2}}\ket{g}+\sin{\frac{\theta}{2}}\ket{e}$. Here the ‘f0g1’ pulse has a peak-truncated Gaussian envelope and a duration of 1000 ns. (b) The reconstructed density matrix of a typical qutrit state, showing a fidelity of 69.19$\% \pm$0.93$\%$. The hollow caps indicate the corresponding ideal density matrix.}
\label{W}
\end{figure}

Propagating photonic $d$-level qudits ($d>2$) can be effectively synthesized with the single-photon source. As high dimensional carriers of quantum information, photonic qudits offer advantages including the storage of exponentially
greater information, larger channel capacity for quantum communication, and simultaneous generation of multiple entangled pairs~\cite{White2008,Siddiqi2021,Jianwei2022,Johannes2022}. In the experiment, we showcase the generation of photonic qutrits ($d=3$) by sequentially transferring the transmon coherence to the propagating photonic mode. 
The pulse sequence is shown in Fig.~\ref{W}(a). We first prepare the transmon qubit to a superposition state $\cos\frac{\theta}{2}\ket{e}+\sin\frac{\theta}{2}\ket{f}$, then apply the `f0g1' pulse to convert the $\ket{f}$ state population to propagating photonic state. After that, we transfer part of the $\ket{e}$ state population to $\ket{f}$ and convert it to the second photonic mode. At last, we apply a $\pi$ pulse of $\ket{e} - \ket{f}$ transition and the `f0g1' pulse to convert the residual excitation to the third photonic mode, resulting in disentangled qubit and photonic states. The final photonic state can be written as $\Psi= \cos\frac{\theta}{2}\cos\frac{\phi}{2}\ket{1_c0_b0_a}+\cos\frac{\theta}{2}\sin\frac{\phi}{2}\ket{0_c1_b0_a}+\sin\frac{\theta}{2}\ket{0_c0_b1_a}$, which is an arbitrary superposition state of a qutrit. By taking $\phi=\pi/2$ and $\theta = \arctan{2\sqrt{2}}$, we have $\Psi = (\ket{1_c0_b0_a}+\ket{0_c1_b0_a}+\ket{0_c0_b1_a})/\sqrt{3}$. In Fig.~\ref{W}(b), we reconstruct the density matrix for this state, obtaining a fidelity of 69.19$\% \pm$0.93$\%$. The state preparation errors are mainly induced by the qubit decay and dephasing during the pulse sequence. Further improvements in the state preparation fidelity can be realized by improving the qubit coherence and reducing the length of the `f0g1' pulse. We note that by sequentially applying `f0g1' transition and properly adjusting the state of the superconducting qubit before each `f0g1' pulse, such a scheme is scalable for the generation of arbitrary qudit state.

\section*{Conclusion}

In conclusion, we realize a frequency-tunable quantum light source in the microwave regime. By embedding a SQUID into the cavity and using the resonator-assisted Raman transition, the quantum state defined in the superconducting qubit can be effectively transferred to a frequency-tunable single-photon state. Hereafter, we showcase the generation of single photons, time-bin encoded propagating photonic qubits and qutrits. The frequency of the emitted photons can be tuned by more than 200 MHz, with a sacrifice of internal quantum efficiency. However, we show that with a proper encoding scheme, the fidelity of the emitted quantum state can be well preserved. As such, our work demonstrates a quantum light source in the microwave regime which has broad application in the future quantum network and modular quantum computing.

\bibliography{qlightref}

\begin{thebibliography}{39}%
\makeatletter
\providecommand \@ifxundefined [1]{%
 \@ifx{#1\undefined}
}%
\providecommand \@ifnum [1]{%
 \ifnum #1\expandafter \@firstoftwo
 \else \expandafter \@secondoftwo
 \fi
}%
\providecommand \@ifx [1]{%
 \ifx #1\expandafter \@firstoftwo
 \else \expandafter \@secondoftwo
 \fi
}%
\providecommand \natexlab [1]{#1}%
\providecommand \enquote  [1]{``#1''}%
\providecommand \bibnamefont  [1]{#1}%
\providecommand \bibfnamefont [1]{#1}%
\providecommand \citenamefont [1]{#1}%
\providecommand \href@noop [0]{\@secondoftwo}%
\providecommand \href [0]{\begingroup \@sanitize@url \@href}%
\providecommand \@href[1]{\@@startlink{#1}\@@href}%
\providecommand \@@href[1]{\endgroup#1\@@endlink}%
\providecommand \@sanitize@url [0]{\catcode `\\12\catcode `\$12\catcode
  `\&12\catcode `\#12\catcode `\^12\catcode `\_12\catcode `\%12\relax}%
\providecommand \@@startlink[1]{}%
\providecommand \@@endlink[0]{}%
\providecommand \url  [0]{\begingroup\@sanitize@url \@url }%
\providecommand \@url [1]{\endgroup\@href {#1}{\urlprefix }}%
\providecommand \urlprefix  [0]{URL }%
\providecommand \Eprint [0]{\href }%
\providecommand \doibase [0]{https://doi.org/}%
\providecommand \selectlanguage [0]{\@gobble}%
\providecommand \bibinfo  [0]{\@secondoftwo}%
\providecommand \bibfield  [0]{\@secondoftwo}%
\providecommand \translation [1]{[#1]}%
\providecommand \BibitemOpen [0]{}%
\providecommand \bibitemStop [0]{}%
\providecommand \bibitemNoStop [0]{.\EOS\space}%
\providecommand \EOS [0]{\spacefactor3000\relax}%
\providecommand \BibitemShut  [1]{\csname bibitem#1\endcsname}%
\let\auto@bib@innerbib\@empty
\bibitem [{\citenamefont {Fabre}\ and\ \citenamefont
  {Treps}(2020)}]{fabre_modes_2020}%
  \BibitemOpen
  \bibfield  {author} {\bibinfo {author} {\bibfnamefont {C.}~\bibnamefont
  {Fabre}}\ and\ \bibinfo {author} {\bibfnamefont {N.}~\bibnamefont {Treps}},\
  }\bibfield  {title} {\bibinfo {title} {Modes and states in quantum optics},\
  }\href {https://doi.org/10.1103/RevModPhys.92.035005} {\bibfield  {journal}
  {\bibinfo  {journal} {Reviews of Modern Physics}\ }\textbf {\bibinfo {volume}
  {92}},\ \bibinfo {pages} {035005} (\bibinfo {year} {2020})},\ \bibinfo {note}
  {publisher: American Physical Society}\BibitemShut {NoStop}%
\bibitem [{\citenamefont {Gu}\ \emph {et~al.}(2017)\citenamefont {Gu},
  \citenamefont {Kockum}, \citenamefont {Miranowicz}, \citenamefont {xi~Liu},\
  and\ \citenamefont {Nori}}]{gu_microwave_2017}%
  \BibitemOpen
  \bibfield  {author} {\bibinfo {author} {\bibfnamefont {X.}~\bibnamefont
  {Gu}}, \bibinfo {author} {\bibfnamefont {A.~F.}\ \bibnamefont {Kockum}},
  \bibinfo {author} {\bibfnamefont {A.}~\bibnamefont {Miranowicz}}, \bibinfo
  {author} {\bibfnamefont {Y.}~\bibnamefont {xi~Liu}},\ and\ \bibinfo {author}
  {\bibfnamefont {F.}~\bibnamefont {Nori}},\ }\bibfield  {title} {\bibinfo
  {title} {Microwave photonics with superconducting quantum circuits},\ }\href
  {https://doi.org/10.1016/j.physrep.2017.10.002} {\bibfield  {journal}
  {\bibinfo  {journal} {Physics Reports}\ }\textbf {\bibinfo {volume}
  {718-719}},\ \bibinfo {pages} {1} (\bibinfo {year} {2017})}\BibitemShut
  {NoStop}%
\bibitem [{\citenamefont {Kimble}(2008)}]{kimble_quantum_2008}%
  \BibitemOpen
  \bibfield  {author} {\bibinfo {author} {\bibfnamefont {H.~J.}\ \bibnamefont
  {Kimble}},\ }\bibfield  {title} {\bibinfo {title} {The quantum internet},\
  }\href {https://doi.org/10.1038/nature07127} {\bibfield  {journal} {\bibinfo
  {journal} {Nature}\ }\textbf {\bibinfo {volume} {453}},\ \bibinfo {pages}
  {1023} (\bibinfo {year} {2008})}\BibitemShut {NoStop}%
\bibitem [{\citenamefont {Bennett}\ and\ \citenamefont
  {Brassard}(1984)}]{BB84}%
  \BibitemOpen
  \bibfield  {author} {\bibinfo {author} {\bibfnamefont {C.~H.}\ \bibnamefont
  {Bennett}}\ and\ \bibinfo {author} {\bibfnamefont {G.}~\bibnamefont
  {Brassard}},\ }\bibfield  {title} {\bibinfo {title} {{Quantum cryptography:
  Public key distribution and coin tossing}},\ }in\ \href@noop {} {\emph
  {\bibinfo {booktitle} {Proceedings of IEEE International Conference on
  Computers, Systems, and Signal Processing}}}\ (\bibinfo {address} {India},\
  \bibinfo {year} {1984})\ p.\ \bibinfo {pages} {175}\BibitemShut {NoStop}%
\bibitem [{\citenamefont {Xu}\ \emph {et~al.}(2020)\citenamefont {Xu},
  \citenamefont {Ma}, \citenamefont {Zhang}, \citenamefont {Lo},\ and\
  \citenamefont {Pan}}]{RevModPhys.92.025002}%
  \BibitemOpen
  \bibfield  {author} {\bibinfo {author} {\bibfnamefont {F.}~\bibnamefont
  {Xu}}, \bibinfo {author} {\bibfnamefont {X.}~\bibnamefont {Ma}}, \bibinfo
  {author} {\bibfnamefont {Q.}~\bibnamefont {Zhang}}, \bibinfo {author}
  {\bibfnamefont {H.-K.}\ \bibnamefont {Lo}},\ and\ \bibinfo {author}
  {\bibfnamefont {J.-W.}\ \bibnamefont {Pan}},\ }\bibfield  {title} {\bibinfo
  {title} {Secure quantum key distribution with realistic devices},\ }\href
  {https://doi.org/10.1103/RevModPhys.92.025002} {\bibfield  {journal}
  {\bibinfo  {journal} {Rev. Mod. Phys.}\ }\textbf {\bibinfo {volume} {92}},\
  \bibinfo {pages} {025002} (\bibinfo {year} {2020})}\BibitemShut {NoStop}%
\bibitem [{\citenamefont {Reiserer}\ and\ \citenamefont
  {Rempe}(2015)}]{reiserer_cavity_based_2015}%
  \BibitemOpen
  \bibfield  {author} {\bibinfo {author} {\bibfnamefont {A.}~\bibnamefont
  {Reiserer}}\ and\ \bibinfo {author} {\bibfnamefont {G.}~\bibnamefont
  {Rempe}},\ }\bibfield  {title} {\bibinfo {title} {Cavity-based quantum
  networks with single atoms and optical photons},\ }\href@noop {} {\bibfield
  {journal} {\bibinfo  {journal} {Rev. Mod. Phys.}\ }\textbf {\bibinfo {volume}
  {87}},\ \bibinfo {pages} {1379} (\bibinfo {year} {2015})}\BibitemShut
  {NoStop}%
\bibitem [{\citenamefont {Duan}\ and\ \citenamefont
  {Monroe}(2010)}]{Monroe2010}%
  \BibitemOpen
  \bibfield  {author} {\bibinfo {author} {\bibfnamefont {L.-M.}\ \bibnamefont
  {Duan}}\ and\ \bibinfo {author} {\bibfnamefont {C.}~\bibnamefont {Monroe}},\
  }\bibfield  {title} {\bibinfo {title} {Colloquium: Quantum networks with
  trapped ions},\ }\href@noop {} {\bibfield  {journal} {\bibinfo  {journal}
  {Rev. Mod. Phys.}\ }\textbf {\bibinfo {volume} {82}},\ \bibinfo {pages}
  {1209} (\bibinfo {year} {2010})}\BibitemShut {NoStop}%
\bibitem [{\citenamefont {Bienfait}\ \emph {et~al.}(2017)\citenamefont
  {Bienfait}, \citenamefont {Campagne-Ibarcq}, \citenamefont {Kiilerich},
  \citenamefont {Zhou}, \citenamefont {Probst}, \citenamefont {Pla},
  \citenamefont {Schenkel}, \citenamefont {Vion}, \citenamefont {Esteve},
  \citenamefont {Morton}, \citenamefont {Moelmer},\ and\ \citenamefont
  {Bertet}}]{bienfait_magnetic_2017}%
  \BibitemOpen
  \bibfield  {author} {\bibinfo {author} {\bibfnamefont {A.}~\bibnamefont
  {Bienfait}}, \bibinfo {author} {\bibfnamefont {P.}~\bibnamefont
  {Campagne-Ibarcq}}, \bibinfo {author} {\bibfnamefont {A.~H.}\ \bibnamefont
  {Kiilerich}}, \bibinfo {author} {\bibfnamefont {X.}~\bibnamefont {Zhou}},
  \bibinfo {author} {\bibfnamefont {S.}~\bibnamefont {Probst}}, \bibinfo
  {author} {\bibfnamefont {J.~J.}\ \bibnamefont {Pla}}, \bibinfo {author}
  {\bibfnamefont {T.}~\bibnamefont {Schenkel}}, \bibinfo {author}
  {\bibfnamefont {D.}~\bibnamefont {Vion}}, \bibinfo {author} {\bibfnamefont
  {D.}~\bibnamefont {Esteve}}, \bibinfo {author} {\bibfnamefont {J.~J.~L.}\
  \bibnamefont {Morton}}, \bibinfo {author} {\bibfnamefont {K.}~\bibnamefont
  {Moelmer}},\ and\ \bibinfo {author} {\bibfnamefont {P.}~\bibnamefont
  {Bertet}},\ }\bibfield  {title} {\bibinfo {title} {Magnetic resonance with
  squeezed microwaves},\ }\href {https://doi.org/10.1103/PhysRevX.7.041011}
  {\bibfield  {journal} {\bibinfo  {journal} {Phys. Rev. X}\ }\textbf {\bibinfo
  {volume} {7}},\ \bibinfo {pages} {041011} (\bibinfo {year}
  {2017})}\BibitemShut {NoStop}%
\bibitem [{\citenamefont {Braunstein}\ and\ \citenamefont {van
  Loock}(2005)}]{braunstein_quantum_2005}%
  \BibitemOpen
  \bibfield  {author} {\bibinfo {author} {\bibfnamefont {S.~L.}\ \bibnamefont
  {Braunstein}}\ and\ \bibinfo {author} {\bibfnamefont {P.}~\bibnamefont {van
  Loock}},\ }\bibfield  {title} {\bibinfo {title} {Quantum information with
  continuous variables},\ }\href {https://doi.org/10.1103/RevModPhys.77.513}
  {\bibfield  {journal} {\bibinfo  {journal} {Rev. Mod. Phys.}\ }\textbf
  {\bibinfo {volume} {77}},\ \bibinfo {pages} {513} (\bibinfo {year}
  {2005})}\BibitemShut {NoStop}%
\bibitem [{\citenamefont {Asavanant}\ \emph {et~al.}(2019)\citenamefont
  {Asavanant}, \citenamefont {Shiozawa}, \citenamefont {Yokoyama},
  \citenamefont {Charoensombutamon}, \citenamefont {Emura}, \citenamefont
  {Alexander}, \citenamefont {Takeda}, \citenamefont {Yoshikawa}, \citenamefont
  {Menicucci}, \citenamefont {Yonezawa},\ and\ \citenamefont
  {Furusawa}}]{asavanant_generation_2019}%
  \BibitemOpen
  \bibfield  {author} {\bibinfo {author} {\bibfnamefont {W.}~\bibnamefont
  {Asavanant}}, \bibinfo {author} {\bibfnamefont {Y.}~\bibnamefont {Shiozawa}},
  \bibinfo {author} {\bibfnamefont {S.}~\bibnamefont {Yokoyama}}, \bibinfo
  {author} {\bibfnamefont {B.}~\bibnamefont {Charoensombutamon}}, \bibinfo
  {author} {\bibfnamefont {H.}~\bibnamefont {Emura}}, \bibinfo {author}
  {\bibfnamefont {R.}~\bibnamefont {Alexander}}, \bibinfo {author}
  {\bibfnamefont {S.}~\bibnamefont {Takeda}}, \bibinfo {author} {\bibfnamefont
  {J.-i.}\ \bibnamefont {Yoshikawa}}, \bibinfo {author} {\bibfnamefont
  {N.}~\bibnamefont {Menicucci}}, \bibinfo {author} {\bibfnamefont
  {H.}~\bibnamefont {Yonezawa}},\ and\ \bibinfo {author} {\bibfnamefont
  {A.}~\bibnamefont {Furusawa}},\ }\bibfield  {title} {\bibinfo {title}
  {Generation of time-domain-multiplexed two-dimensional cluster state},\
  }\href {https://doi.org/10.1126/science.aay2645} {\bibfield  {journal}
  {\bibinfo  {journal} {Science}\ }\textbf {\bibinfo {volume} {366}},\ \bibinfo
  {pages} {373} (\bibinfo {year} {2019})}\BibitemShut {NoStop}%
\bibitem [{\citenamefont {Larsen}\ \emph {et~al.}(2019)\citenamefont {Larsen},
  \citenamefont {Guo}, \citenamefont {Breum}, \citenamefont
  {Neergaard-Nielsen},\ and\ \citenamefont
  {Andersen}}]{larsen_deterministic_2019}%
  \BibitemOpen
  \bibfield  {author} {\bibinfo {author} {\bibfnamefont {M.}~\bibnamefont
  {Larsen}}, \bibinfo {author} {\bibfnamefont {X.}~\bibnamefont {Guo}},
  \bibinfo {author} {\bibfnamefont {C.}~\bibnamefont {Breum}}, \bibinfo
  {author} {\bibfnamefont {J.}~\bibnamefont {Neergaard-Nielsen}},\ and\
  \bibinfo {author} {\bibfnamefont {U.}~\bibnamefont {Andersen}},\ }\bibfield
  {title} {\bibinfo {title} {Deterministic generation of a two-dimensional
  cluster state},\ }\href {https://doi.org/10.1126/science.aay4354} {\bibfield
  {journal} {\bibinfo  {journal} {Science}\ }\textbf {\bibinfo {volume}
  {366}},\ \bibinfo {pages} {369} (\bibinfo {year} {2019})}\BibitemShut
  {NoStop}%
\bibitem [{\citenamefont {Cirac}\ \emph {et~al.}(1997)\citenamefont {Cirac},
  \citenamefont {Zoller}, \citenamefont {Kimble},\ and\ \citenamefont
  {Mabuchi}}]{Cirac_97_transfer}%
  \BibitemOpen
  \bibfield  {author} {\bibinfo {author} {\bibfnamefont {J.~I.}\ \bibnamefont
  {Cirac}}, \bibinfo {author} {\bibfnamefont {P.}~\bibnamefont {Zoller}},
  \bibinfo {author} {\bibfnamefont {H.~J.}\ \bibnamefont {Kimble}},\ and\
  \bibinfo {author} {\bibfnamefont {H.}~\bibnamefont {Mabuchi}},\ }\bibfield
  {title} {\bibinfo {title} {Quantum state transfer and entanglement
  distribution among distant nodes in a quantum network},\ }\href
  {https://doi.org/10.1103/PhysRevLett.78.3221} {\bibfield  {journal} {\bibinfo
   {journal} {Phys. Rev. Lett.}\ }\textbf {\bibinfo {volume} {78}},\ \bibinfo
  {pages} {3221} (\bibinfo {year} {1997})}\BibitemShut {NoStop}%
\bibitem [{\citenamefont {Narla}\ \emph {et~al.}(2016)\citenamefont {Narla},
  \citenamefont {Shankar}, \citenamefont {Hatridge}, \citenamefont {Leghtas},
  \citenamefont {Sliwa}, \citenamefont {Zalys-Geller}, \citenamefont
  {Mundhada}, \citenamefont {Pfaff}, \citenamefont {Frunzio}, \citenamefont
  {Schoelkopf},\ and\ \citenamefont {Devoret}}]{Devoret2016}%
  \BibitemOpen
  \bibfield  {author} {\bibinfo {author} {\bibfnamefont {A.}~\bibnamefont
  {Narla}}, \bibinfo {author} {\bibfnamefont {S.}~\bibnamefont {Shankar}},
  \bibinfo {author} {\bibfnamefont {M.}~\bibnamefont {Hatridge}}, \bibinfo
  {author} {\bibfnamefont {Z.}~\bibnamefont {Leghtas}}, \bibinfo {author}
  {\bibfnamefont {K.~M.}\ \bibnamefont {Sliwa}}, \bibinfo {author}
  {\bibfnamefont {E.}~\bibnamefont {Zalys-Geller}}, \bibinfo {author}
  {\bibfnamefont {S.~O.}\ \bibnamefont {Mundhada}}, \bibinfo {author}
  {\bibfnamefont {W.}~\bibnamefont {Pfaff}}, \bibinfo {author} {\bibfnamefont
  {L.}~\bibnamefont {Frunzio}}, \bibinfo {author} {\bibfnamefont {R.~J.}\
  \bibnamefont {Schoelkopf}},\ and\ \bibinfo {author} {\bibfnamefont {M.~H.}\
  \bibnamefont {Devoret}},\ }\bibfield  {title} {\bibinfo {title} {Robust
  concurrent remote entanglement between two superconducting qubits},\ }\href
  {https://doi.org/10.1103/PhysRevX.6.031036} {\bibfield  {journal} {\bibinfo
  {journal} {Phys. Rev. X}\ }\textbf {\bibinfo {volume} {6}},\ \bibinfo {pages}
  {031036} (\bibinfo {year} {2016})}\BibitemShut {NoStop}%
\bibitem [{\citenamefont {Kurpiers}\ \emph {et~al.}(2018)\citenamefont
  {Kurpiers}, \citenamefont {Magnard}, \citenamefont {Walter}, \citenamefont
  {Royer}, \citenamefont {Pechal}, \citenamefont {Heinsoo}, \citenamefont
  {Salath{\'{e}}}, \citenamefont {Akin}, \citenamefont {Storz}, \citenamefont
  {Besse}, \citenamefont {Gasparinetti}, \citenamefont {Blais},\ and\
  \citenamefont {Wallraff}}]{Kurpiers_2018}%
  \BibitemOpen
  \bibfield  {author} {\bibinfo {author} {\bibfnamefont {P.}~\bibnamefont
  {Kurpiers}}, \bibinfo {author} {\bibfnamefont {P.}~\bibnamefont {Magnard}},
  \bibinfo {author} {\bibfnamefont {T.}~\bibnamefont {Walter}}, \bibinfo
  {author} {\bibfnamefont {B.}~\bibnamefont {Royer}}, \bibinfo {author}
  {\bibfnamefont {M.}~\bibnamefont {Pechal}}, \bibinfo {author} {\bibfnamefont
  {J.}~\bibnamefont {Heinsoo}}, \bibinfo {author} {\bibfnamefont
  {Y.}~\bibnamefont {Salath{\'{e}}}}, \bibinfo {author} {\bibfnamefont
  {A.}~\bibnamefont {Akin}}, \bibinfo {author} {\bibfnamefont {S.}~\bibnamefont
  {Storz}}, \bibinfo {author} {\bibfnamefont {J.-C.}\ \bibnamefont {Besse}},
  \bibinfo {author} {\bibfnamefont {S.}~\bibnamefont {Gasparinetti}}, \bibinfo
  {author} {\bibfnamefont {A.}~\bibnamefont {Blais}},\ and\ \bibinfo {author}
  {\bibfnamefont {A.}~\bibnamefont {Wallraff}},\ }\bibfield  {title} {\bibinfo
  {title} {Deterministic quantum state transfer and remote entanglement using
  microwave photons},\ }\href {https://doi.org/10.1038/s41586-018-0195-y}
  {\bibfield  {journal} {\bibinfo  {journal} {Nature}\ }\textbf {\bibinfo
  {volume} {558}},\ \bibinfo {pages} {264} (\bibinfo {year}
  {2018})}\BibitemShut {NoStop}%
\bibitem [{\citenamefont {Kurpiers}\ \emph {et~al.}(2019)\citenamefont
  {Kurpiers}, \citenamefont {Pechal}, \citenamefont {Royer}, \citenamefont
  {Magnard}, \citenamefont {Walter}, \citenamefont {Heinsoo}, \citenamefont
  {Salath\'e}, \citenamefont {Akin}, \citenamefont {Storz}, \citenamefont
  {Besse}, \citenamefont {Gasparinetti}, \citenamefont {Blais},\ and\
  \citenamefont {Wallraff}}]{Kurpiers_2019}%
  \BibitemOpen
  \bibfield  {author} {\bibinfo {author} {\bibfnamefont {P.}~\bibnamefont
  {Kurpiers}}, \bibinfo {author} {\bibfnamefont {M.}~\bibnamefont {Pechal}},
  \bibinfo {author} {\bibfnamefont {B.}~\bibnamefont {Royer}}, \bibinfo
  {author} {\bibfnamefont {P.}~\bibnamefont {Magnard}}, \bibinfo {author}
  {\bibfnamefont {T.}~\bibnamefont {Walter}}, \bibinfo {author} {\bibfnamefont
  {J.}~\bibnamefont {Heinsoo}}, \bibinfo {author} {\bibfnamefont
  {Y.}~\bibnamefont {Salath\'e}}, \bibinfo {author} {\bibfnamefont
  {A.}~\bibnamefont {Akin}}, \bibinfo {author} {\bibfnamefont {S.}~\bibnamefont
  {Storz}}, \bibinfo {author} {\bibfnamefont {J.-C.}\ \bibnamefont {Besse}},
  \bibinfo {author} {\bibfnamefont {S.}~\bibnamefont {Gasparinetti}}, \bibinfo
  {author} {\bibfnamefont {A.}~\bibnamefont {Blais}},\ and\ \bibinfo {author}
  {\bibfnamefont {A.}~\bibnamefont {Wallraff}},\ }\bibfield  {title} {\bibinfo
  {title} {Quantum communication with time-bin encoded microwave photons},\
  }\href@noop {} {\bibfield  {journal} {\bibinfo  {journal} {Phys. Rev. Appl.}\
  }\textbf {\bibinfo {volume} {12}},\ \bibinfo {pages} {044067} (\bibinfo
  {year} {2019})}\BibitemShut {NoStop}%
\bibitem [{\citenamefont {Zhong}\ \emph {et~al.}(2021)\citenamefont {Zhong},
  \citenamefont {Chang}, \citenamefont {Bienfait}, \citenamefont {Dumur},
  \citenamefont {Chou}, \citenamefont {Conner}, \citenamefont {Grebel},
  \citenamefont {Povey}, \citenamefont {Yan}, \citenamefont {Schuster},\ and\
  \citenamefont {Cleland}}]{zhong2021}%
  \BibitemOpen
  \bibfield  {author} {\bibinfo {author} {\bibfnamefont {Y.}~\bibnamefont
  {Zhong}}, \bibinfo {author} {\bibfnamefont {H.-S.}\ \bibnamefont {Chang}},
  \bibinfo {author} {\bibfnamefont {A.}~\bibnamefont {Bienfait}}, \bibinfo
  {author} {\bibfnamefont {{\'{E}}.}~\bibnamefont {Dumur}}, \bibinfo {author}
  {\bibfnamefont {M.-H.}\ \bibnamefont {Chou}}, \bibinfo {author}
  {\bibfnamefont {C.~R.}\ \bibnamefont {Conner}}, \bibinfo {author}
  {\bibfnamefont {J.}~\bibnamefont {Grebel}}, \bibinfo {author} {\bibfnamefont
  {R.~G.}\ \bibnamefont {Povey}}, \bibinfo {author} {\bibfnamefont
  {H.}~\bibnamefont {Yan}}, \bibinfo {author} {\bibfnamefont {D.~I.}\
  \bibnamefont {Schuster}},\ and\ \bibinfo {author} {\bibfnamefont {A.~N.}\
  \bibnamefont {Cleland}},\ }\bibfield  {title} {\bibinfo {title}
  {Deterministic multi-qubit entanglement in a quantum network},\ }\href
  {https://doi.org/10.1038/s41586-021-03288-7} {\bibfield  {journal} {\bibinfo
  {journal} {Nature}\ }\textbf {\bibinfo {volume} {590}},\ \bibinfo {pages}
  {571} (\bibinfo {year} {2021})}\BibitemShut {NoStop}%
\bibitem [{\citenamefont {Qiu}\ \emph {et~al.}(2023)\citenamefont {Qiu},
  \citenamefont {Liu}, \citenamefont {Niu}, \citenamefont {Hu}, \citenamefont
  {Wu}, \citenamefont {Zhang}, \citenamefont {Huang}, \citenamefont {Chen},
  \citenamefont {Li}, \citenamefont {Liu}, \citenamefont {Zhong}, \citenamefont
  {Duan},\ and\ \citenamefont {Yu}}]{zhong2023}%
  \BibitemOpen
  \bibfield  {author} {\bibinfo {author} {\bibfnamefont {J.}~\bibnamefont
  {Qiu}}, \bibinfo {author} {\bibfnamefont {Y.}~\bibnamefont {Liu}}, \bibinfo
  {author} {\bibfnamefont {J.}~\bibnamefont {Niu}}, \bibinfo {author}
  {\bibfnamefont {L.}~\bibnamefont {Hu}}, \bibinfo {author} {\bibfnamefont
  {Y.}~\bibnamefont {Wu}}, \bibinfo {author} {\bibfnamefont {L.}~\bibnamefont
  {Zhang}}, \bibinfo {author} {\bibfnamefont {W.}~\bibnamefont {Huang}},
  \bibinfo {author} {\bibfnamefont {Y.}~\bibnamefont {Chen}}, \bibinfo {author}
  {\bibfnamefont {J.}~\bibnamefont {Li}}, \bibinfo {author} {\bibfnamefont
  {S.}~\bibnamefont {Liu}}, \bibinfo {author} {\bibfnamefont {Y.}~\bibnamefont
  {Zhong}}, \bibinfo {author} {\bibfnamefont {L.}~\bibnamefont {Duan}},\ and\
  \bibinfo {author} {\bibfnamefont {D.}~\bibnamefont {Yu}},\ }\href
  {https://doi.org/10.48550/ARXIV.2302.08756} {\bibinfo {title} {Deterministic
  quantum teleportation between distant superconducting chips}} (\bibinfo
  {year} {2023})\BibitemShut {NoStop}%
\bibitem [{\citenamefont {Shields}(2007)}]{shields_semiconductor_2007}%
  \BibitemOpen
  \bibfield  {author} {\bibinfo {author} {\bibfnamefont {A.~J.}\ \bibnamefont
  {Shields}},\ }\bibfield  {title} {\bibinfo {title} {Semiconductor quantum
  light sources},\ }\href {https://doi.org/10.1038/nphoton.2007.46} {\bibfield
  {journal} {\bibinfo  {journal} {Nature Photonics}\ }\textbf {\bibinfo
  {volume} {1}},\ \bibinfo {pages} {215} (\bibinfo {year} {2007})}\BibitemShut
  {NoStop}%
\bibitem [{\citenamefont {Senellart}\ \emph {et~al.}(2017)\citenamefont
  {Senellart}, \citenamefont {Solomon},\ and\ \citenamefont
  {White}}]{senellart_high_performance_2017}%
  \BibitemOpen
  \bibfield  {author} {\bibinfo {author} {\bibfnamefont {P.}~\bibnamefont
  {Senellart}}, \bibinfo {author} {\bibfnamefont {G.}~\bibnamefont {Solomon}},\
  and\ \bibinfo {author} {\bibfnamefont {A.}~\bibnamefont {White}},\ }\bibfield
   {title} {\bibinfo {title} {High-performance semiconductor quantum-dot
  single-photon sources},\ }\href {https://doi.org/10.1038/nnano.2017.218}
  {\bibfield  {journal} {\bibinfo  {journal} {Nature Nanotechnology}\ }\textbf
  {\bibinfo {volume} {12}},\ \bibinfo {pages} {1026} (\bibinfo {year}
  {2017})}\BibitemShut {NoStop}%
\bibitem [{\citenamefont {Thomas}\ \emph {et~al.}(2022)\citenamefont {Thomas},
  \citenamefont {Ruscio}, \citenamefont {Morin},\ and\ \citenamefont
  {Rempe}}]{Thomas_rempe_2022}%
  \BibitemOpen
  \bibfield  {author} {\bibinfo {author} {\bibfnamefont {P.}~\bibnamefont
  {Thomas}}, \bibinfo {author} {\bibfnamefont {L.}~\bibnamefont {Ruscio}},
  \bibinfo {author} {\bibfnamefont {O.}~\bibnamefont {Morin}},\ and\ \bibinfo
  {author} {\bibfnamefont {G.}~\bibnamefont {Rempe}},\ }\bibfield  {title}
  {\bibinfo {title} {Efficient generation of entangled multiphoton graph states
  from a single atom},\ }\href {https://doi.org/10.1038/s41586-022-04987-5}
  {\bibfield  {journal} {\bibinfo  {journal} {Nature}\ }\textbf {\bibinfo
  {volume} {608}},\ \bibinfo {pages} {677} (\bibinfo {year}
  {2022})}\BibitemShut {NoStop}%
\bibitem [{\citenamefont {Blais}\ \emph {et~al.}(2021)\citenamefont {Blais},
  \citenamefont {Grimsmo}, \citenamefont {Girvin},\ and\ \citenamefont
  {Wallraff}}]{blais_circuit_2021}%
  \BibitemOpen
  \bibfield  {author} {\bibinfo {author} {\bibfnamefont {A.}~\bibnamefont
  {Blais}}, \bibinfo {author} {\bibfnamefont {A.~L.}\ \bibnamefont {Grimsmo}},
  \bibinfo {author} {\bibfnamefont {S.~M.}\ \bibnamefont {Girvin}},\ and\
  \bibinfo {author} {\bibfnamefont {A.}~\bibnamefont {Wallraff}},\ }\bibfield
  {title} {\bibinfo {title} {Circuit quantum electrodynamics},\ }\href
  {https://doi.org/10.1103/RevModPhys.93.025005} {\bibfield  {journal}
  {\bibinfo  {journal} {Rev. Mod. Phys.}\ }\textbf {\bibinfo {volume} {93}},\
  \bibinfo {pages} {025005} (\bibinfo {year} {2021})}\BibitemShut {NoStop}%
\bibitem [{\citenamefont {Bao}\ \emph {et~al.}(2022)\citenamefont {Bao},
  \citenamefont {Wang}, \citenamefont {Wu}, \citenamefont {Li}, \citenamefont
  {Cai}, \citenamefont {Wang}, \citenamefont {Ma}, \citenamefont {Cai},
  \citenamefont {Han}, \citenamefont {Wang}, \citenamefont {Song},
  \citenamefont {Sun}, \citenamefont {Zhang},\ and\ \citenamefont
  {Duan}}]{Bao_2022}%
  \BibitemOpen
  \bibfield  {author} {\bibinfo {author} {\bibfnamefont {Z.}~\bibnamefont
  {Bao}}, \bibinfo {author} {\bibfnamefont {Z.}~\bibnamefont {Wang}}, \bibinfo
  {author} {\bibfnamefont {Y.}~\bibnamefont {Wu}}, \bibinfo {author}
  {\bibfnamefont {Y.}~\bibnamefont {Li}}, \bibinfo {author} {\bibfnamefont
  {W.}~\bibnamefont {Cai}}, \bibinfo {author} {\bibfnamefont {W.}~\bibnamefont
  {Wang}}, \bibinfo {author} {\bibfnamefont {Y.}~\bibnamefont {Ma}}, \bibinfo
  {author} {\bibfnamefont {T.}~\bibnamefont {Cai}}, \bibinfo {author}
  {\bibfnamefont {X.}~\bibnamefont {Han}}, \bibinfo {author} {\bibfnamefont
  {J.}~\bibnamefont {Wang}}, \bibinfo {author} {\bibfnamefont {Y.}~\bibnamefont
  {Song}}, \bibinfo {author} {\bibfnamefont {L.}~\bibnamefont {Sun}}, \bibinfo
  {author} {\bibfnamefont {H.}~\bibnamefont {Zhang}},\ and\ \bibinfo {author}
  {\bibfnamefont {L.}~\bibnamefont {Duan}},\ }\bibfield  {title} {\bibinfo
  {title} {Experimental preparation of generalized cat states for itinerant
  microwave photons},\ }\href@noop {} {\bibfield  {journal} {\bibinfo
  {journal} {Phys. Rev. A}\ }\textbf {\bibinfo {volume} {105}},\ \bibinfo
  {pages} {063717} (\bibinfo {year} {2022})}\BibitemShut {NoStop}%
\bibitem [{\citenamefont {Wang}\ \emph {et~al.}(2022)\citenamefont {Wang},
  \citenamefont {Bao}, \citenamefont {Wu}, \citenamefont {Li}, \citenamefont
  {Cai}, \citenamefont {Wang}, \citenamefont {Ma}, \citenamefont {Cai},
  \citenamefont {Han}, \citenamefont {Wang}, \citenamefont {Song},
  \citenamefont {Sun}, \citenamefont {Zhang},\ and\ \citenamefont
  {Duan}}]{Wang_2022}%
  \BibitemOpen
  \bibfield  {author} {\bibinfo {author} {\bibfnamefont {Z.}~\bibnamefont
  {Wang}}, \bibinfo {author} {\bibfnamefont {Z.}~\bibnamefont {Bao}}, \bibinfo
  {author} {\bibfnamefont {Y.}~\bibnamefont {Wu}}, \bibinfo {author}
  {\bibfnamefont {Y.}~\bibnamefont {Li}}, \bibinfo {author} {\bibfnamefont
  {W.}~\bibnamefont {Cai}}, \bibinfo {author} {\bibfnamefont {W.}~\bibnamefont
  {Wang}}, \bibinfo {author} {\bibfnamefont {Y.}~\bibnamefont {Ma}}, \bibinfo
  {author} {\bibfnamefont {T.}~\bibnamefont {Cai}}, \bibinfo {author}
  {\bibfnamefont {X.}~\bibnamefont {Han}}, \bibinfo {author} {\bibfnamefont
  {J.}~\bibnamefont {Wang}}, \bibinfo {author} {\bibfnamefont {Y.}~\bibnamefont
  {Song}}, \bibinfo {author} {\bibfnamefont {L.}~\bibnamefont {Sun}}, \bibinfo
  {author} {\bibfnamefont {H.}~\bibnamefont {Zhang}},\ and\ \bibinfo {author}
  {\bibfnamefont {L.}~\bibnamefont {Duan}},\ }\bibfield  {title} {\bibinfo
  {title} {A flying schrödinger's cat in multipartite entangled states},\
  }\href@noop {} {\bibfield  {journal} {\bibinfo  {journal} {Science Advances}\
  }\textbf {\bibinfo {volume} {8}} (\bibinfo {year} {2022})}\BibitemShut
  {NoStop}%
\bibitem [{\citenamefont {Eichler}\ \emph
  {et~al.}(2011{\natexlab{a}})\citenamefont {Eichler}, \citenamefont
  {Bozyigit}, \citenamefont {Lang}, \citenamefont {Steffen}, \citenamefont
  {Fink},\ and\ \citenamefont {Wallraff}}]{eichler_experimental_2011}%
  \BibitemOpen
  \bibfield  {author} {\bibinfo {author} {\bibfnamefont {C.}~\bibnamefont
  {Eichler}}, \bibinfo {author} {\bibfnamefont {D.}~\bibnamefont {Bozyigit}},
  \bibinfo {author} {\bibfnamefont {C.}~\bibnamefont {Lang}}, \bibinfo {author}
  {\bibfnamefont {L.}~\bibnamefont {Steffen}}, \bibinfo {author} {\bibfnamefont
  {J.}~\bibnamefont {Fink}},\ and\ \bibinfo {author} {\bibfnamefont
  {A.}~\bibnamefont {Wallraff}},\ }\bibfield  {title} {\bibinfo {title}
  {Experimental state tomography of itinerant single microwave photons},\
  }\href {https://doi.org/10.1103/PhysRevLett.106.220503} {\bibfield  {journal}
  {\bibinfo  {journal} {Phys. Rev. Lett.}\ }\textbf {\bibinfo {volume} {106}},\
  \bibinfo {pages} {220503} (\bibinfo {year} {2011}{\natexlab{a}})}\BibitemShut
  {NoStop}%
\bibitem [{\citenamefont {Lang}\ \emph {et~al.}(2013)\citenamefont {Lang},
  \citenamefont {Eichler}, \citenamefont {Steffen}, \citenamefont {Fink},
  \citenamefont {Woolley}, \citenamefont {Blais},\ and\ \citenamefont
  {Wallraff}}]{lang_correlations_2013}%
  \BibitemOpen
  \bibfield  {author} {\bibinfo {author} {\bibfnamefont {C.}~\bibnamefont
  {Lang}}, \bibinfo {author} {\bibfnamefont {C.}~\bibnamefont {Eichler}},
  \bibinfo {author} {\bibfnamefont {L.}~\bibnamefont {Steffen}}, \bibinfo
  {author} {\bibfnamefont {J.}~\bibnamefont {Fink}}, \bibinfo {author}
  {\bibfnamefont {M.}~\bibnamefont {Woolley}}, \bibinfo {author} {\bibfnamefont
  {A.}~\bibnamefont {Blais}},\ and\ \bibinfo {author} {\bibfnamefont
  {A.}~\bibnamefont {Wallraff}},\ }\bibfield  {title} {\bibinfo {title}
  {Correlations, indistinguishability and entanglement in hong-ou-mandel
  experiments at microwave frequencies},\ }\href
  {https://doi.org/10.1038/nphys2612} {\bibfield  {journal} {\bibinfo
  {journal} {Nature Physics}\ }\textbf {\bibinfo {volume} {9}},\ \bibinfo
  {pages} {345} (\bibinfo {year} {2013})}\BibitemShut {NoStop}%
\bibitem [{\citenamefont {Besse}\ \emph {et~al.}(2020)\citenamefont {Besse},
  \citenamefont {Reuer}, \citenamefont {Collodo}, \citenamefont {Wulff},
  \citenamefont {Wernli}, \citenamefont {Copetudo}, \citenamefont {Malz},
  \citenamefont {Magnard}, \citenamefont {Akin}, \citenamefont {Gabureac},
  \citenamefont {Norris}, \citenamefont {Cirac}, \citenamefont {Wallraff},\
  and\ \citenamefont {Eichler}}]{besse_realizing_2020}%
  \BibitemOpen
  \bibfield  {author} {\bibinfo {author} {\bibfnamefont {J.-C.}\ \bibnamefont
  {Besse}}, \bibinfo {author} {\bibfnamefont {K.}~\bibnamefont {Reuer}},
  \bibinfo {author} {\bibfnamefont {M.~C.}\ \bibnamefont {Collodo}}, \bibinfo
  {author} {\bibfnamefont {A.}~\bibnamefont {Wulff}}, \bibinfo {author}
  {\bibfnamefont {L.}~\bibnamefont {Wernli}}, \bibinfo {author} {\bibfnamefont
  {A.}~\bibnamefont {Copetudo}}, \bibinfo {author} {\bibfnamefont
  {D.}~\bibnamefont {Malz}}, \bibinfo {author} {\bibfnamefont {P.}~\bibnamefont
  {Magnard}}, \bibinfo {author} {\bibfnamefont {A.}~\bibnamefont {Akin}},
  \bibinfo {author} {\bibfnamefont {M.}~\bibnamefont {Gabureac}}, \bibinfo
  {author} {\bibfnamefont {G.~J.}\ \bibnamefont {Norris}}, \bibinfo {author}
  {\bibfnamefont {J.~I.}\ \bibnamefont {Cirac}}, \bibinfo {author}
  {\bibfnamefont {A.}~\bibnamefont {Wallraff}},\ and\ \bibinfo {author}
  {\bibfnamefont {C.}~\bibnamefont {Eichler}},\ }\bibfield  {title} {\bibinfo
  {title} {Realizing a deterministic source of multipartite-entangled photonic
  qubits},\ }\href@noop {} {\bibfield  {journal} {\bibinfo  {journal} {Nature
  Communications}\ }\textbf {\bibinfo {volume} {11}} (\bibinfo {year}
  {2020})}\BibitemShut {NoStop}%
\bibitem [{\citenamefont {Pechal}\ \emph {et~al.}(2014)\citenamefont {Pechal},
  \citenamefont {Huthmacher}, \citenamefont {Eichler}, \citenamefont
  {Zeytino\ifmmode~\breve{g}\else \u{g}\fi{}lu}, \citenamefont {Abdumalikov},
  \citenamefont {Berger}, \citenamefont {Wallraff},\ and\ \citenamefont
  {Filipp}}]{pechal_microwave_controlled_2014}%
  \BibitemOpen
  \bibfield  {author} {\bibinfo {author} {\bibfnamefont {M.}~\bibnamefont
  {Pechal}}, \bibinfo {author} {\bibfnamefont {L.}~\bibnamefont {Huthmacher}},
  \bibinfo {author} {\bibfnamefont {C.}~\bibnamefont {Eichler}}, \bibinfo
  {author} {\bibfnamefont {S.}~\bibnamefont {Zeytino\ifmmode~\breve{g}\else
  \u{g}\fi{}lu}}, \bibinfo {author} {\bibfnamefont {A.~A.}\ \bibnamefont
  {Abdumalikov}}, \bibinfo {author} {\bibfnamefont {S.}~\bibnamefont {Berger}},
  \bibinfo {author} {\bibfnamefont {A.}~\bibnamefont {Wallraff}},\ and\
  \bibinfo {author} {\bibfnamefont {S.}~\bibnamefont {Filipp}},\ }\bibfield
  {title} {\bibinfo {title} {Microwave-controlled generation of shaped single
  photons in circuit quantum electrodynamics},\ }\href
  {https://doi.org/10.1103/PhysRevX.4.041010} {\bibfield  {journal} {\bibinfo
  {journal} {Phys. Rev. X}\ }\textbf {\bibinfo {volume} {4}},\ \bibinfo {pages}
  {041010} (\bibinfo {year} {2014})}\BibitemShut {NoStop}%
\bibitem [{\citenamefont {Ilves}\ \emph {et~al.}(2020)\citenamefont {Ilves},
  \citenamefont {Kono}, \citenamefont {Sunada}, \citenamefont {Yamazaki},
  \citenamefont {Kim}, \citenamefont {Koshino},\ and\ \citenamefont
  {Nakamura}}]{ilves_demand_2020}%
  \BibitemOpen
  \bibfield  {author} {\bibinfo {author} {\bibfnamefont {J.}~\bibnamefont
  {Ilves}}, \bibinfo {author} {\bibfnamefont {S.}~\bibnamefont {Kono}},
  \bibinfo {author} {\bibfnamefont {Y.}~\bibnamefont {Sunada}}, \bibinfo
  {author} {\bibfnamefont {S.}~\bibnamefont {Yamazaki}}, \bibinfo {author}
  {\bibfnamefont {M.}~\bibnamefont {Kim}}, \bibinfo {author} {\bibfnamefont
  {K.}~\bibnamefont {Koshino}},\ and\ \bibinfo {author} {\bibfnamefont
  {Y.}~\bibnamefont {Nakamura}},\ }\bibfield  {title} {\bibinfo {title}
  {On-demand generation and characterization of a microwave time-bin qubit},\
  }\href@noop {} {\bibfield  {journal} {\bibinfo  {journal} {npj Quantum
  Information}\ }\textbf {\bibinfo {volume} {6}} (\bibinfo {year}
  {2020})}\BibitemShut {NoStop}%
\bibitem [{\citenamefont {Zhu}\ \emph {et~al.}(2022)\citenamefont {Zhu},
  \citenamefont {Chen}, \citenamefont {Yu}, \citenamefont {Shao}, \citenamefont
  {Hu}, \citenamefont {Xin}, \citenamefont {Yeh}, \citenamefont {Ghosh},
  \citenamefont {He}, \citenamefont {Reimer}, \citenamefont {Sinclair},
  \citenamefont {Wong}, \citenamefont {Zhang},\ and\ \citenamefont
  {Loncar}}]{zhu_spectral_2022}%
  \BibitemOpen
  \bibfield  {author} {\bibinfo {author} {\bibfnamefont {D.}~\bibnamefont
  {Zhu}}, \bibinfo {author} {\bibfnamefont {C.}~\bibnamefont {Chen}}, \bibinfo
  {author} {\bibfnamefont {M.}~\bibnamefont {Yu}}, \bibinfo {author}
  {\bibfnamefont {L.}~\bibnamefont {Shao}}, \bibinfo {author} {\bibfnamefont
  {Y.}~\bibnamefont {Hu}}, \bibinfo {author} {\bibfnamefont {C.}~\bibnamefont
  {Xin}}, \bibinfo {author} {\bibfnamefont {M.}~\bibnamefont {Yeh}}, \bibinfo
  {author} {\bibfnamefont {S.}~\bibnamefont {Ghosh}}, \bibinfo {author}
  {\bibfnamefont {L.}~\bibnamefont {He}}, \bibinfo {author} {\bibfnamefont
  {C.}~\bibnamefont {Reimer}}, \bibinfo {author} {\bibfnamefont
  {N.}~\bibnamefont {Sinclair}}, \bibinfo {author} {\bibfnamefont
  {F.}~\bibnamefont {Wong}}, \bibinfo {author} {\bibfnamefont {M.}~\bibnamefont
  {Zhang}},\ and\ \bibinfo {author} {\bibfnamefont {M.}~\bibnamefont
  {Loncar}},\ }\bibfield  {title} {\bibinfo {title} {Spectral control of
  nonclassical light pulses using an integrated thin-film lithium niobate
  modulator},\ }\href {https://doi.org/10.1038/s41377-022-01029-7} {\bibfield
  {journal} {\bibinfo  {journal} {Light: Science Applications}\ }\textbf
  {\bibinfo {volume} {11}},\ \bibinfo {pages} {327} (\bibinfo {year}
  {2022})}\BibitemShut {NoStop}%
\bibitem [{\citenamefont {Kiilerich}\ and\ \citenamefont
  {M\o{}lmer}(2019)}]{Klaus_2019}%
  \BibitemOpen
  \bibfield  {author} {\bibinfo {author} {\bibfnamefont {A.~H.}\ \bibnamefont
  {Kiilerich}}\ and\ \bibinfo {author} {\bibfnamefont {K.}~\bibnamefont
  {M\o{}lmer}},\ }\bibfield  {title} {\bibinfo {title} {Input-output theory
  with quantum pulses},\ }\href
  {https://doi.org/10.1103/PhysRevLett.123.123604} {\bibfield  {journal}
  {\bibinfo  {journal} {Phys. Rev. Lett.}\ }\textbf {\bibinfo {volume} {123}},\
  \bibinfo {pages} {123604} (\bibinfo {year} {2019})}\BibitemShut {NoStop}%
\bibitem [{\citenamefont {Bao}\ \emph {et~al.}(2021)\citenamefont {Bao},
  \citenamefont {Wang}, \citenamefont {Wu}, \citenamefont {Li}, \citenamefont
  {Ma}, \citenamefont {Song}, \citenamefont {Zhang},\ and\ \citenamefont
  {Duan}}]{Bao_memory_2021}%
  \BibitemOpen
  \bibfield  {author} {\bibinfo {author} {\bibfnamefont {Z.}~\bibnamefont
  {Bao}}, \bibinfo {author} {\bibfnamefont {Z.}~\bibnamefont {Wang}}, \bibinfo
  {author} {\bibfnamefont {Y.}~\bibnamefont {Wu}}, \bibinfo {author}
  {\bibfnamefont {Y.}~\bibnamefont {Li}}, \bibinfo {author} {\bibfnamefont
  {C.}~\bibnamefont {Ma}}, \bibinfo {author} {\bibfnamefont {Y.}~\bibnamefont
  {Song}}, \bibinfo {author} {\bibfnamefont {H.}~\bibnamefont {Zhang}},\ and\
  \bibinfo {author} {\bibfnamefont {L.}~\bibnamefont {Duan}},\ }\bibfield
  {title} {\bibinfo {title} {On-demand storage and retrieval of microwave
  photons using a superconducting multiresonator quantum memory},\ }\href
  {https://doi.org/10.1103/PhysRevLett.127.010503} {\bibfield  {journal}
  {\bibinfo  {journal} {Phys. Rev. Lett.}\ }\textbf {\bibinfo {volume} {127}},\
  \bibinfo {pages} {010503} (\bibinfo {year} {2021})}\BibitemShut {NoStop}%
\bibitem [{\citenamefont {Eichler}\ \emph
  {et~al.}(2011{\natexlab{b}})\citenamefont {Eichler}, \citenamefont
  {Bozyigit}, \citenamefont {Lang}, \citenamefont {Steffen}, \citenamefont
  {Fink},\ and\ \citenamefont {Wallraff}}]{Wallraff2011}%
  \BibitemOpen
  \bibfield  {author} {\bibinfo {author} {\bibfnamefont {C.}~\bibnamefont
  {Eichler}}, \bibinfo {author} {\bibfnamefont {D.}~\bibnamefont {Bozyigit}},
  \bibinfo {author} {\bibfnamefont {C.}~\bibnamefont {Lang}}, \bibinfo {author}
  {\bibfnamefont {L.}~\bibnamefont {Steffen}}, \bibinfo {author} {\bibfnamefont
  {J.}~\bibnamefont {Fink}},\ and\ \bibinfo {author} {\bibfnamefont
  {A.}~\bibnamefont {Wallraff}},\ }\bibfield  {title} {\bibinfo {title}
  {Experimental state tomography of itinerant single microwave photons},\
  }\href@noop {} {\bibfield  {journal} {\bibinfo  {journal} {Phys. Rev. Lett.}\
  }\textbf {\bibinfo {volume} {106}},\ \bibinfo {pages} {220503} (\bibinfo
  {year} {2011}{\natexlab{b}})}\BibitemShut {NoStop}%
\bibitem [{\citenamefont {Sandberg}\ \emph {et~al.}(2008)\citenamefont
  {Sandberg}, \citenamefont {Wilson}, \citenamefont {Persson}, \citenamefont
  {Bauch}, \citenamefont {Johansson}, \citenamefont {Shumeiko}, \citenamefont
  {Duty},\ and\ \citenamefont {Delsing}}]{Sandberg_Tuning_2008}%
  \BibitemOpen
  \bibfield  {author} {\bibinfo {author} {\bibfnamefont {M.}~\bibnamefont
  {Sandberg}}, \bibinfo {author} {\bibfnamefont {C.}~\bibnamefont {Wilson}},
  \bibinfo {author} {\bibfnamefont {F.}~\bibnamefont {Persson}}, \bibinfo
  {author} {\bibfnamefont {T.}~\bibnamefont {Bauch}}, \bibinfo {author}
  {\bibfnamefont {G.}~\bibnamefont {Johansson}}, \bibinfo {author}
  {\bibfnamefont {V.}~\bibnamefont {Shumeiko}}, \bibinfo {author}
  {\bibfnamefont {T.}~\bibnamefont {Duty}},\ and\ \bibinfo {author}
  {\bibfnamefont {P.}~\bibnamefont {Delsing}},\ }\bibfield  {title} {\bibinfo
  {title} {Tuning the field in a microwave resonator faster than the photon
  lifetime},\ }\href {https://doi.org/10.1063/1.2929367} {\bibfield  {journal}
  {\bibinfo  {journal} {Applied Physics Letters}\ }\textbf {\bibinfo {volume}
  {92}},\ \bibinfo {pages} {203501} (\bibinfo {year} {2008})}\BibitemShut
  {NoStop}%
\bibitem [{\citenamefont {Eichler}\ \emph {et~al.}(2012)\citenamefont
  {Eichler}, \citenamefont {Bozyigit},\ and\ \citenamefont
  {Wallraff}}]{Eichler_Characterizing_2012}%
  \BibitemOpen
  \bibfield  {author} {\bibinfo {author} {\bibfnamefont {C.}~\bibnamefont
  {Eichler}}, \bibinfo {author} {\bibfnamefont {D.}~\bibnamefont {Bozyigit}},\
  and\ \bibinfo {author} {\bibfnamefont {A.}~\bibnamefont {Wallraff}},\
  }\bibfield  {title} {\bibinfo {title} {Characterizing quantum microwave
  radiation and its entanglement with superconducting qubits using linear
  detectors},\ }\href {https://doi.org/10.1103/PhysRevA.86.032106} {\bibfield
  {journal} {\bibinfo  {journal} {Phys. Rev. A}\ }\textbf {\bibinfo {volume}
  {86}},\ \bibinfo {pages} {032106} (\bibinfo {year} {2012})}\BibitemShut
  {NoStop}%
\bibitem [{\citenamefont {Knall}\ \emph {et~al.}(2022)\citenamefont {Knall},
  \citenamefont {Knaut}, \citenamefont {Bekenstein}, \citenamefont {Assumpcao},
  \citenamefont {Stroganov}, \citenamefont {Gong}, \citenamefont {Huan},
  \citenamefont {Stas}, \citenamefont {Machielse}, \citenamefont {Chalupnik},
  \citenamefont {Levonian}, \citenamefont {Suleymanzade}, \citenamefont
  {Riedinger}, \citenamefont {Park}, \citenamefont {Lon\ifmmode~\check{c}\else
  \v{c}\fi{}ar}, \citenamefont {Bhaskar},\ and\ \citenamefont
  {Lukin}}]{Lukin2022}%
  \BibitemOpen
  \bibfield  {author} {\bibinfo {author} {\bibfnamefont {E.~N.}\ \bibnamefont
  {Knall}}, \bibinfo {author} {\bibfnamefont {C.~M.}\ \bibnamefont {Knaut}},
  \bibinfo {author} {\bibfnamefont {R.}~\bibnamefont {Bekenstein}}, \bibinfo
  {author} {\bibfnamefont {D.~R.}\ \bibnamefont {Assumpcao}}, \bibinfo {author}
  {\bibfnamefont {P.~L.}\ \bibnamefont {Stroganov}}, \bibinfo {author}
  {\bibfnamefont {W.}~\bibnamefont {Gong}}, \bibinfo {author} {\bibfnamefont
  {Y.~Q.}\ \bibnamefont {Huan}}, \bibinfo {author} {\bibfnamefont {P.-J.}\
  \bibnamefont {Stas}}, \bibinfo {author} {\bibfnamefont {B.}~\bibnamefont
  {Machielse}}, \bibinfo {author} {\bibfnamefont {M.}~\bibnamefont
  {Chalupnik}}, \bibinfo {author} {\bibfnamefont {D.}~\bibnamefont {Levonian}},
  \bibinfo {author} {\bibfnamefont {A.}~\bibnamefont {Suleymanzade}}, \bibinfo
  {author} {\bibfnamefont {R.}~\bibnamefont {Riedinger}}, \bibinfo {author}
  {\bibfnamefont {H.}~\bibnamefont {Park}}, \bibinfo {author} {\bibfnamefont
  {M.}~\bibnamefont {Lon\ifmmode~\check{c}\else \v{c}\fi{}ar}}, \bibinfo
  {author} {\bibfnamefont {M.~K.}\ \bibnamefont {Bhaskar}},\ and\ \bibinfo
  {author} {\bibfnamefont {M.~D.}\ \bibnamefont {Lukin}},\ }\bibfield  {title}
  {\bibinfo {title} {Efficient source of shaped single photons based on an
  integrated diamond nanophotonic system},\ }\href
  {https://doi.org/10.1103/PhysRevLett.129.053603} {\bibfield  {journal}
  {\bibinfo  {journal} {Phys. Rev. Lett.}\ }\textbf {\bibinfo {volume} {129}},\
  \bibinfo {pages} {053603} (\bibinfo {year} {2022})}\BibitemShut {NoStop}%
\bibitem [{\citenamefont {Lanyon}\ \emph {et~al.}(2008)\citenamefont {Lanyon},
  \citenamefont {Weinhold}, \citenamefont {Langford}, \citenamefont {O'Brien},
  \citenamefont {Resch}, \citenamefont {Gilchrist},\ and\ \citenamefont
  {White}}]{White2008}%
  \BibitemOpen
  \bibfield  {author} {\bibinfo {author} {\bibfnamefont {B.~P.}\ \bibnamefont
  {Lanyon}}, \bibinfo {author} {\bibfnamefont {T.~J.}\ \bibnamefont
  {Weinhold}}, \bibinfo {author} {\bibfnamefont {N.~K.}\ \bibnamefont
  {Langford}}, \bibinfo {author} {\bibfnamefont {J.~L.}\ \bibnamefont
  {O'Brien}}, \bibinfo {author} {\bibfnamefont {K.~J.}\ \bibnamefont {Resch}},
  \bibinfo {author} {\bibfnamefont {A.}~\bibnamefont {Gilchrist}},\ and\
  \bibinfo {author} {\bibfnamefont {A.~G.}\ \bibnamefont {White}},\ }\bibfield
  {title} {\bibinfo {title} {Manipulating biphotonic qutrits},\ }\href
  {https://doi.org/10.1103/PhysRevLett.100.060504} {\bibfield  {journal}
  {\bibinfo  {journal} {Phys. Rev. Lett.}\ }\textbf {\bibinfo {volume} {100}},\
  \bibinfo {pages} {060504} (\bibinfo {year} {2008})}\BibitemShut {NoStop}%
\bibitem [{\citenamefont {Morvan}\ \emph {et~al.}(2021)\citenamefont {Morvan},
  \citenamefont {Ramasesh}, \citenamefont {Blok}, \citenamefont {Kreikebaum},
  \citenamefont {O'Brien}, \citenamefont {Chen}, \citenamefont {Mitchell},
  \citenamefont {Naik}, \citenamefont {Santiago},\ and\ \citenamefont
  {Siddiqi}}]{Siddiqi2021}%
  \BibitemOpen
  \bibfield  {author} {\bibinfo {author} {\bibfnamefont {A.}~\bibnamefont
  {Morvan}}, \bibinfo {author} {\bibfnamefont {V.~V.}\ \bibnamefont
  {Ramasesh}}, \bibinfo {author} {\bibfnamefont {M.~S.}\ \bibnamefont {Blok}},
  \bibinfo {author} {\bibfnamefont {J.~M.}\ \bibnamefont {Kreikebaum}},
  \bibinfo {author} {\bibfnamefont {K.}~\bibnamefont {O'Brien}}, \bibinfo
  {author} {\bibfnamefont {L.}~\bibnamefont {Chen}}, \bibinfo {author}
  {\bibfnamefont {B.~K.}\ \bibnamefont {Mitchell}}, \bibinfo {author}
  {\bibfnamefont {R.~K.}\ \bibnamefont {Naik}}, \bibinfo {author}
  {\bibfnamefont {D.~I.}\ \bibnamefont {Santiago}},\ and\ \bibinfo {author}
  {\bibfnamefont {I.}~\bibnamefont {Siddiqi}},\ }\bibfield  {title} {\bibinfo
  {title} {Qutrit randomized benchmarking},\ }\href
  {https://doi.org/10.1103/PhysRevLett.126.210504} {\bibfield  {journal}
  {\bibinfo  {journal} {Phys. Rev. Lett.}\ }\textbf {\bibinfo {volume} {126}},\
  \bibinfo {pages} {210504} (\bibinfo {year} {2021})}\BibitemShut {NoStop}%
\bibitem [{\citenamefont {Chi}\ \emph {et~al.}(2022)\citenamefont {Chi},
  \citenamefont {Huang}, \citenamefont {Zhang}, \citenamefont {Mao},
  \citenamefont {Zhou}, \citenamefont {Chen}, \citenamefont {Zhai},
  \citenamefont {Bao}, \citenamefont {Dai}, \citenamefont {Yuan}, \citenamefont
  {Zhang}, \citenamefont {Dai}, \citenamefont {Tang}, \citenamefont {Yang},
  \citenamefont {Li}, \citenamefont {Ding}, \citenamefont {Oxenløwe},
  \citenamefont {Thompson}, \citenamefont {O'Brien},\ and\ \citenamefont
  {Wang}}]{Jianwei2022}%
  \BibitemOpen
  \bibfield  {author} {\bibinfo {author} {\bibfnamefont {Y.}~\bibnamefont
  {Chi}}, \bibinfo {author} {\bibfnamefont {J.}~\bibnamefont {Huang}}, \bibinfo
  {author} {\bibfnamefont {Z.}~\bibnamefont {Zhang}}, \bibinfo {author}
  {\bibfnamefont {J.}~\bibnamefont {Mao}}, \bibinfo {author} {\bibfnamefont
  {Z.}~\bibnamefont {Zhou}}, \bibinfo {author} {\bibfnamefont {X.}~\bibnamefont
  {Chen}}, \bibinfo {author} {\bibfnamefont {C.}~\bibnamefont {Zhai}}, \bibinfo
  {author} {\bibfnamefont {J.}~\bibnamefont {Bao}}, \bibinfo {author}
  {\bibfnamefont {T.}~\bibnamefont {Dai}}, \bibinfo {author} {\bibfnamefont
  {H.}~\bibnamefont {Yuan}}, \bibinfo {author} {\bibfnamefont {M.}~\bibnamefont
  {Zhang}}, \bibinfo {author} {\bibfnamefont {D.}~\bibnamefont {Dai}}, \bibinfo
  {author} {\bibfnamefont {B.}~\bibnamefont {Tang}}, \bibinfo {author}
  {\bibfnamefont {Y.}~\bibnamefont {Yang}}, \bibinfo {author} {\bibfnamefont
  {Z.}~\bibnamefont {Li}}, \bibinfo {author} {\bibfnamefont {Y.}~\bibnamefont
  {Ding}}, \bibinfo {author} {\bibfnamefont {L.}~\bibnamefont {Oxenløwe}},
  \bibinfo {author} {\bibfnamefont {M.}~\bibnamefont {Thompson}}, \bibinfo
  {author} {\bibfnamefont {J.}~\bibnamefont {O'Brien}},\ and\ \bibinfo {author}
  {\bibfnamefont {J.}~\bibnamefont {Wang}},\ }\bibfield  {title} {\bibinfo
  {title} {A programmable qudit-based quantum processor},\ }\href
  {https://doi.org/10.1038/s41467-022-28767-x} {\bibfield  {journal} {\bibinfo
  {journal} {Nature Communications}\ }\textbf {\bibinfo {volume} {13}},\
  \bibinfo {pages} {1166} (\bibinfo {year} {2022})}\BibitemShut {NoStop}%
\bibitem [{\citenamefont {Zheng}\ \emph {et~al.}(2022)\citenamefont {Zheng},
  \citenamefont {Sharma},\ and\ \citenamefont {Borregaard}}]{Johannes2022}%
  \BibitemOpen
  \bibfield  {author} {\bibinfo {author} {\bibfnamefont {Y.}~\bibnamefont
  {Zheng}}, \bibinfo {author} {\bibfnamefont {H.}~\bibnamefont {Sharma}},\ and\
  \bibinfo {author} {\bibfnamefont {J.}~\bibnamefont {Borregaard}},\ }\bibfield
   {title} {\bibinfo {title} {Entanglement distribution with minimal memory
  requirements using time-bin photonic qudits},\ }\href
  {https://doi.org/10.1103/PRXQuantum.3.040319} {\bibfield  {journal} {\bibinfo
   {journal} {PRX Quantum}\ }\textbf {\bibinfo {volume} {3}},\ \bibinfo {pages}
  {040319} (\bibinfo {year} {2022})}\BibitemShut {NoStop}%
\end{thebibliography}%

\bigskip

\textbf{Acknowledgments:} This work was supported by Innovation Program for Quantum Science and Technology (2021ZD0301704), the Tsinghua University Initiative Scientific Research Program, and the Ministry of Education of China.

\textbf{Competing interests:} The authors declare that there are no competing interests.

\textbf{Data Availability:} The data that support the findings of this study are available from the authors upon request.

\end{document}